# Structural studies on Multiferroic $La_{1-x}Bi_xCrO_3$ Perovskites

Aga Shahee, Dhirendra Kumar, N. P. Lalla[#]

*UGC-DAE Consortium for Scientific Research, University campus, Khandwa Road, Indore-452001*
*[#]nplalla@csr.res.in*

**Abstract**. The synthesis and structural studies on the polycrystalline $La_{1-x}Bi_xCrO_3$ ( x =0 , 0.2, 0.25, 0.3, 0.35) have been done using XRD and TEM. The results of the Reitveld refinement of XRD data shows that these compounds crystallize in an orthorhombic structure with Pnma space group. With increasing Bi content, the unit cell volume and lattice parameters undergo non monotonous changes within the tolerance factor from 0.84092 to 0.84072. This has been attributed to the anisotropy induced by lone pair of $Bi^{3+}$, which may distorted the lattice.

## INTRODUCTION

Bi-containing perovskites have attracted much attention during the last decade as multiferroic and lead-free ferroelectric materials. The presence of the stereochemically-active lone pair of a $Bi^{3+}$ ion [1] introduces interesting physical properties in Bi doped perovskites. In multiferroic systems, two or more ferroic orders are present in the same phase[1]. The coupling between the magnetic order and ferroelectric order can arise due to magentoelectric (ME) effect [2]. This ME effect provides an additional degree of freedom in the design of actuators, transducers, and next generation memory devices. Like $BiFeO_3$, $BiCrO_3$ is also an important perovskite compound expected to show enhance multiferroic properties [3]. But synthesis of $BiCrO_3$ needs high temperature and high pressure. Therefore we have studied $Bi^{3+}$ doped $LaCrO_3$ which is likely to show multiferroic behaviour and will also help to understand the basic properties of $BiCrO_3$. This paper reports the results obtained on the synthesis, structural and microstructural analysis of solid solutions of LaCrO3 and $BiCrO_3$.

## EXPERMENTAL DETAILS

Perovskite oxides $La_{1-x}Bi_xCrO_3$ (with x=0.0, 0.2, 0.25, 0.3 and 0.35) were prepared by conventional solid-state reaction route using $La_2O_3$ (99.99%), $Cr_2O_3$ (99.99%) and $Bi_2O_3$ (99.99%). Since Bi has high vapour pressure, the samples where prepared with 5% excess of $Bi_2O_3$ to maintain desired stoichiometry of the final product. The stoichiometric mixtures of reactants with excess of $Bi_2O_3$ were thoroughly ground for ~10 hours. Few drops of water solution of polyvinyl alcohol (1wt %) was added as binder during final grinding before pelletization. The pellets were first preheated at 550°C in air for 2h to remove the binder and then fired at 1250°C for 5-8 hours. As sintered pellet were then cut across the thickness into two pieces and subjected to bulk structural characterization using XRD and microstructural characterization, in imaging and diffraction mode, using FEI-TECNAI-20-$G^2$ transmission electron microscope (TEM).

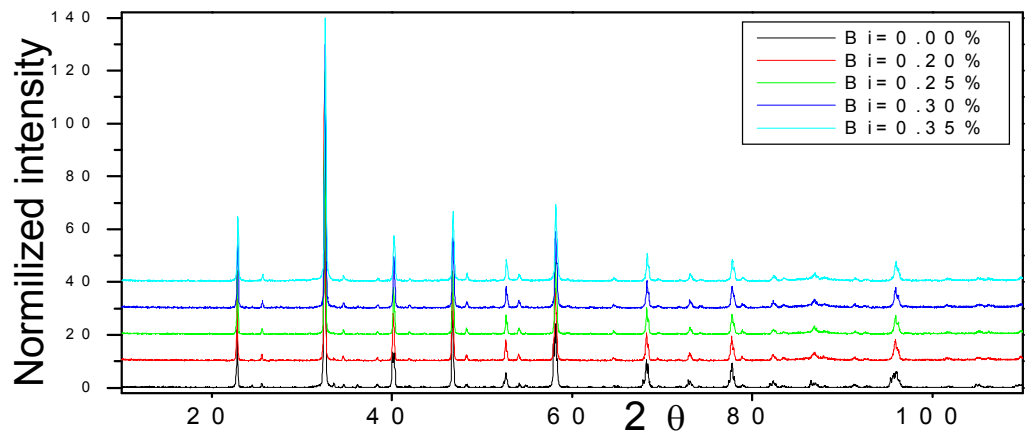

**Fig.1.a.** XRD patterns of $La_{1-x}Bi_xCrO_3$.

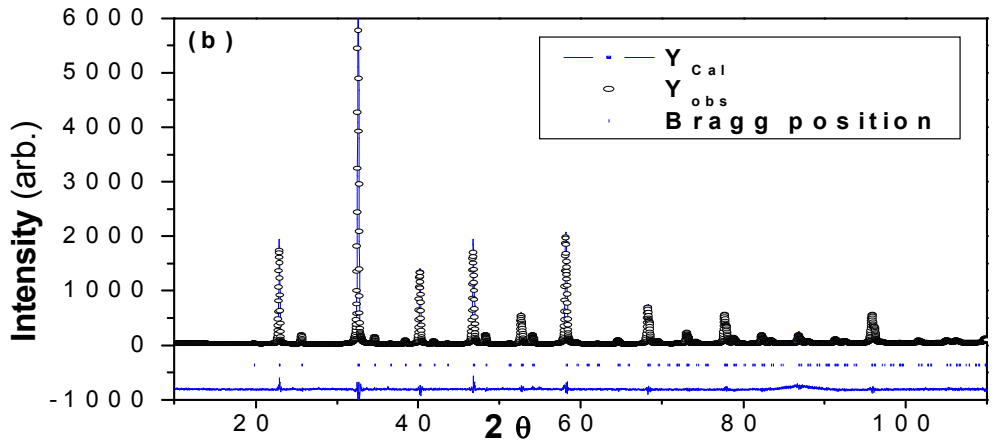

**Fig.1.b.** Reitveld refined XRD pattern of $La_{0.70}Bi_{0.30}CrO_3$.

All the cut pieces were studied using XRD with Cu-Kα in the 2θ range of 10°-110° using Rigaku powder X-ray diffractometer. To investigate the presence of unreacted $Bi_2O_3$, which is quit likely to be deposited on the surface during furnace cooling, XRD data, was collected from both bare upper and inner cut surfaces. In fact we did find unreacted $Bi_2O_3$ on the bare upper surface of the pellets.

## RESULTS AND DISSCUSION

Fig.1 (a) shows the XRD patterns of the inner cut surfaces (bulk) only. The bulk was found free from unreacted $Bi_2O_3$. The XRD data were refinement using the FULL PROOF-2K Rietveld refinement Program. Using this we could obtain accurate lattice parameters and Wyckoff positions as a function of Bi content. Fig.1 (b) represents a typical Rietveld refined XRD data for x=0.3.

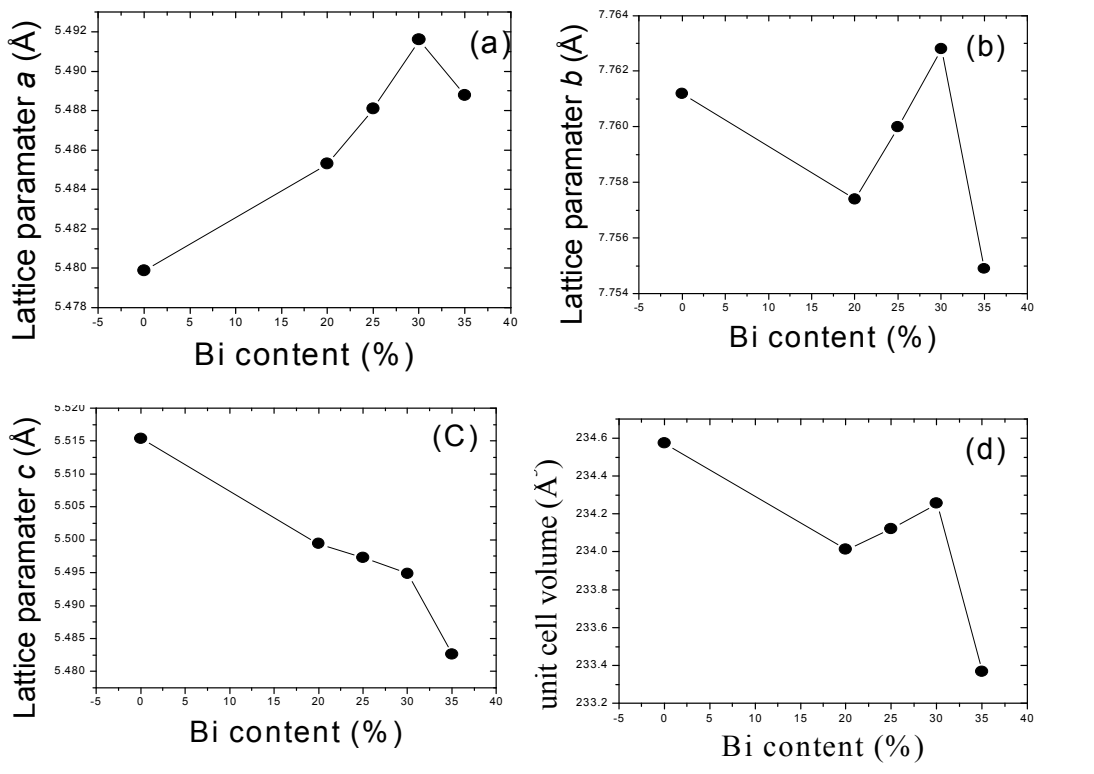

**Fig.2 (a-d).** Variations of unit cell parameters and unit cell volume as a function of Bi content.

Keeping in view that Bi will probably go to La site forming solid solution and also looking at the similarity between the observed XRD patterns and that of the reported $LaCrO_3$[4], an input file was made based on the Wyckoff positions and the starting lattice parameters as reported for $LaCrO_3$[4]. In the input file Bi was kept at La site. The space-group used for refinement was Pnma. All the observed peaks were very nicely accounted by Pnma phase. We could not observe any un accounted peak. This confirms the formation of single phase solid solution of $La_{1-x}Bi_xCrO_3$ with orthorhombicaly distorted perovskite with Pnma space-group. Based on the final refinement results a graph of comparison of lattice parameters and unit cell volume was constructed; see fig.2 (a-d). It can be seen that with increasing $Bi^{3+}$ content the lattice parameters and cell volume undergo non monotonous changes. Up to 30% Bi substitution lattice parameter 'a' increases and at 35% it decreases where as the lattice parameter 'b' and unit cell volume first decreases, up to 20%, then increases up to 30%. Above 30% it abruptly decreases. The 'c' parameter keeps decreasing slowly up to 30% and above 30% this rate increases. Keeping in view the nearly identical ionic radii of $La^{3+}$ and $Bi^{3+}$ (1.32 Å and 1.30Å, respectively) the observed non monotonous changes in the lattice parameters appears to be due to anisotropic distortion in Bi-O bond lengths arising due to $Bi^{3+}$ lone pair. The distortion of Bi-O cause octahedral tilting leading to change in lattice parameters. TEM was carried out on well characterized sample of $La_{0.70}Bi_{0.30}CrO_3$. Fig.3 (a) shows an electron micrograph of densely packed grains of sizes ~1μm. Fig.3 (b, c) represents selected area diffraction (SAD) patterns taken from a single grain of $La_{0.70}Bi_{0.30}CrO_3$. Through measurement it was confirmed that the SADs belong to the orthorhombic phase with lattice parameters a= 5.48Å, b= 7.76 Å & c= 5.49Å and correspond to (b) [010] and (c) [101] zones of the Pnma phase. This is in confirmation with Reitveld refined XRD results.

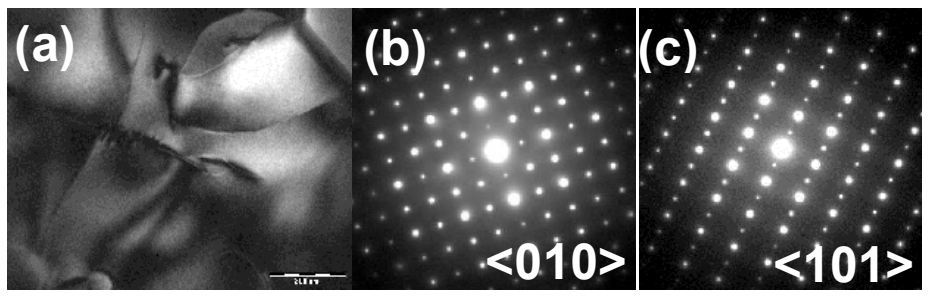

**Fig.3.** Microstructure (a) and SAD patterns taken along (b) [010], and (c) [101] zones of the orthorhombic Pnma phase of $La_{0.70}Bi_{0.30}CrO_3$.

## CONCLUSION

Single phase multiferroic solid solutions of $La_{1-x}Bi_xCrO_3$ have been successfully synthesized employing solid-state reaction in excess of $Bi_2O_3$. These solid solutions crystallize in a orthorhombicaly distorted perovskite phase with space group Pnma**.** With increasing $Bi^{3+}$ content the lattice parameters and the cell volume undergo non monotonous changes. Keeping in view the nearly identical ionic radii of $La^{3+}$ and $Bi^{3+}$ the observed non monotonous changes in the lattice parameters appears to be due to anisotropic distortion in Bi-O bond lengths arising due to anisotropic interaction of $Bi^{3+}$ lone pair electron. The distortion in Bi-O causes octahedral tilting leading to change in lattice parameters.

## ACKNOWLEGEMENT

Aga Shahee would like to acknowledge CSIR-India for financial support.